\documentclass[conference]{IEEEtran}
\IEEEoverridecommandlockouts
\usepackage{cite}
\usepackage{amsmath,amssymb,amsfonts}
\usepackage{algorithmic}
\usepackage{graphicx}
\usepackage{textcomp}
\usepackage{xcolor}
\usepackage{amsfonts}
\usepackage{booktabs}
\usepackage{float}
\usepackage{graphicx}
\usepackage{grffile}
\usepackage[algo2e]{algorithm2e}
\usepackage{longtable}
\usepackage{wrapfig}
\usepackage{gensymb}
\usepackage{textcomp}
\usepackage{amsmath}
\usepackage{algorithm}
\usepackage{verbatim}
\usepackage{graphicx}
\usepackage{caption}
\usepackage{listings}
\usepackage{array}
\usepackage{subcaption}
\usepackage{booktabs}
\usepackage{amsmath}
\usepackage{enumitem}
\usepackage{dblfloatfix}
\usepackage{balance}
\usepackage{lipsum}
\def\arraystretch{1.25}
\usepackage[utf8]{inputenc}

\begin{document}
\title{A Game Theoretic Analysis for \\ Cooperative Smart Farming}
\makeatletter
\newcommand{\linebreakand}{%
  \end{@IEEEauthorhalign}
  \hfill\mbox{}\par
  \mbox{}\hfill\begin{@IEEEauthorhalign}
}
\author{\IEEEauthorblockN{Deepti Gupta\IEEEauthorrefmark{1}\IEEEauthorrefmark{4},  Paras Bhatt\IEEEauthorrefmark{2}\IEEEauthorrefmark{4}, Smriti Bhatt\IEEEauthorrefmark{3}}
\IEEEauthorblockA{\IEEEauthorrefmark{1}Dept. of Computer Science,
University of Texas at San Antonio,
San Antonio, Texas 78249, USA \\
\IEEEauthorrefmark{2}Dept. of Information Systems and Cyber Security,
University of Texas at San Antonio,
San Antonio, Texas 78249, USA\\
\IEEEauthorrefmark{3}Dept. of Computing and Cyber Security, Texas A \& M University-San Antonio,
San Antonio, Texas 78224, USA\\}
\IEEEauthorrefmark{1}deepti.mrt@gmail.com,
\IEEEauthorrefmark{2}paras.bhatt@utsa.edu, 
\IEEEauthorrefmark{3}sbhatt@tamusa.edu \\
\thanks{\IEEEauthorrefmark{4} These authors contributed equally to this work.}}
\maketitle

\begin{abstract}
The application of Internet of Things (IoT) and Machine Learning (ML) to the agricultural industry has enabled the development and creation of smart farms and precision agriculture. The growth in the number of smart farms and potential cooperation between these farms has given rise to the Cooperative Smart Farming (CSF) where different connected farms collaborate with each other and share data for their mutual benefit. This data sharing through CSF has various advantages where individual data from separate farms can be aggregated by ML models and be used to produce actionable outputs which then can be utilized by all the farms in CSFs. This enables farms to gain better insights for enhancing desired outputs, such as crop yield, managing water resources and irrigation schedules, as well as better seed applications. However, complications may arise in CSF when some of the farms do not transfer high-quality data and rather rely on other farms to feed ML models. Another possibility is the presence of rogue farms in CSFs that want to snoop on other farms without actually contributing any data. In this paper, we analyze the behavior of farms participating in CSFs using game theory approach, where each farm is motivated to maximize its profit. We first present the problem of defective farms in CSFs due to lack of better data, and then propose a ML framework that segregates farms and automatically assign them to an appropriate CSF cluster based on the quality of data they provide. Our proposed model rewards the farms supplying better data and penalize the ones that do not provide required data or are malicious in nature, thus, ensuring the model integrity and better performance all over while solving the defective farms problem. 

\end{abstract}
\section{Introduction}
\label{intro}
Farming is one the earliest activities undertaken by humans to build civilizations. Feeding the ever-increasing world population is a huge challenge and the role farming in meeting this challenge cannot be overstated. The recent recognition of the World Food Program (WFP)\footnote{https://www.nobelprize.org/prizes/peace/2020/summary/} as the Nobel Peace Prize recipient underlines the importance food security plays in our lives. Consequently, it is equally important to ensure that sustainable farming practices are followed so that land is not overused and exploited. Internet of Things (IoT) technology can serve as a critical asset to enhance agricultural practices and enable efficient and sustainable farming. With the application of IoT and Machine Learning (ML) algorithms in the farming sector, a new domain of smart agriculture has emerged. Using IoT to monitor and measure different aspects of soil, land and crops can lead to development of better and more sophisticated farms that use data-driven models and applications to enhance their production, which is also known as smart farming and/or precision agriculture today. 

With the rapid diffusion of smart devices and IoT technology in the agriculture domain, farms have grown more advanced, technically competent and production efficient \cite{zhao2010study}. The use of IoT has been abundant and in turn led to the creation of a technology based smart ecosystem. Such a connected system enabled by smart devices and key technologies as cloud and edge computing leads to the creation of smart farms which generates large amounts of data. This large amount of data can be analyzed using ML and Artificial Intelligence (AI) technologies to extract crucial insights and implications. The need for such smart farms has been stressed in research and is equally important for the sustenance of smart cities \cite{hallett2017smart, gupta2020future}. As smart cities develop so should the infrastructure around farming, as it is not only an extension but a necessary derivative of the fully connected IoT architecture. 
However, the increase in number of smart connected farms has not been without issues. There is a possibility of adversaries sabotaging farms, malicious agents compromising the functioning of such farms or even cyber-attacks against smart farming adopters \cite{gupta2020security, chukkapalli2020smart}. It is therefore important to have strong security in place that thwarts such attacks and better data management practices so that informed decisions can be made that do not jeopardize smart farm outputs. Another reason to ensure data security and management is to increase the farm owners’ confidence in the ability of technology to actually provide them with visible benefits. The benefits accruing from data analysis of smart farms can either be related to greater production output, or better land use, or even the discovery of efficient seed application techniques.

The use of cloud computing as well as web-based services has ushered in a new era where smart farms plan better yield management, adopt responsible farming practices, and indulge in sustainable techniques. It is for this reason that IT has been so readily deployed and used in the agriculture sector, giving rise to the smart farms concept \cite{colezea2018cluefarm}. Smart farming involves the adoption of IoT devices to automate processes on regular farms, such as irrigation and seed application. It also involves collecting data on the soil quality, moisture and crop yield metrics. Such data can drive new knowledge discovery which benefits and informs the entire agricultural industry of new and efficient ways to farm and grow crops. Also, as new developments take place in the IoT industry and advances are made in data usage techniques, it is equally important that these improvements rapidly percolate to the agriculture domain primarily to the associated smart farms.

The increasing number of smart farms demands cooperation between each other which allows these farms to learn efficient farming techniques by applying data-driven models to smart farm data and share resources among each other. This type of cooperation between several smart farms is known as Cooperative Smart Farming (CSF). While CSF is an attractive framework to benefit multiple farms together, it is important to understand that there are also security and privacy issues that could arise in this framework. Data privacy could be a major concern for some smart farms, thus, it is essential to address security and privacy in CSFs. There may be different types of farms participating in a CSF. A farm, which has low-quality agriculture data generated from IoT devices, always wants to be a part of CSF to take benefit from other farms. A farm, which has huge number of resources and high-quality agriculture data, does not receive any benefit in CSF. Therefore, there is an issue for smart farms to become a member of CSF and contribute effectively as member of the CSF. In this paper, we address the research problem of unfair collaboration among smart farms in a CSF by proposing a CSF game model, and a novel fair strategy which enforces each smart farm to cooperate in CSF based on the clusters formed to achieve benefit through AI assisted application. The main contributions of this paper are as follows. 
\begin{enumerate}
    \item We identify the issue of unfair collaboration among farms in cooperative smart farming. For instance, some smart farms which generate low-quality data to build machine learning model can take advantage from other farms which have high-quality data.
    \item We present two different use case scenarios including different types of smart farms in a CSF along with rationality assumption.
    \item We propose a fair strategy to enforce farms to cooperate in CSF to build ML model.
    \item We present a proposed implementation framework that can be employed and extended to enable CSF use case scenarios utilizing a real-world cloud-enabled IoT platform, Amazon Web Services (AWS). 
\end{enumerate}

The rest of paper is organized as follows. Section \ref{related} presents relevant work on CSF, security and privacy issues in CSF, and game theoretic models used in smart farming. Cooperative smart farming use case scenarios along with rational assumption are discussed in Section \ref{usecase}. Section \ref{gamemodel} shows cooperative smart farming game and also shows the game analysis in same section. Section \ref{approach} presents proposed fair strategy. Section \ref{result} discuss proposed implementation framework for enabling CSF using real-world cloud-enabled IoT platform. Section \ref{conclusion} concludes the paper with future research directions. 

 \section{Related Work}
 In this section, we discuss the relevant work in perspective of three specific areas - \textit{cooperative smart farming (CSF), security and privacy issues in CSFs,} and \textit{game theory application in the context of smart farming}. 
 \label{related}
 \subsection{Cooperative Smart Farming}
Farming has always been a community activity and in its truest sense a communal one that includes a participatory process. Therefore, even though individual farms operating alone may be efficient revenue maximizers, it is still an evident fact that significant benefits can accrue to those participating in a cooperative bloc \cite{chinn1980cooperative}. Similar to cooperation in the traditional sense of farming, where farmers share resources such as seeds and equipment; a CSF scenario in the technical sense involves sharing of data collected by numerous sensors, devices, and other resources. The reasons to indulge in CSF are not only economic ones but also ones of security. For instance, the participating farms share data which can lead to better crop output. Meanwhile, the data shared can also be analyzed to figure out threats to crops and protect them from damage \cite{satamraju2017rural}.
\begin{figure*}[t]
\centering
\includegraphics[width=0.75\textwidth]{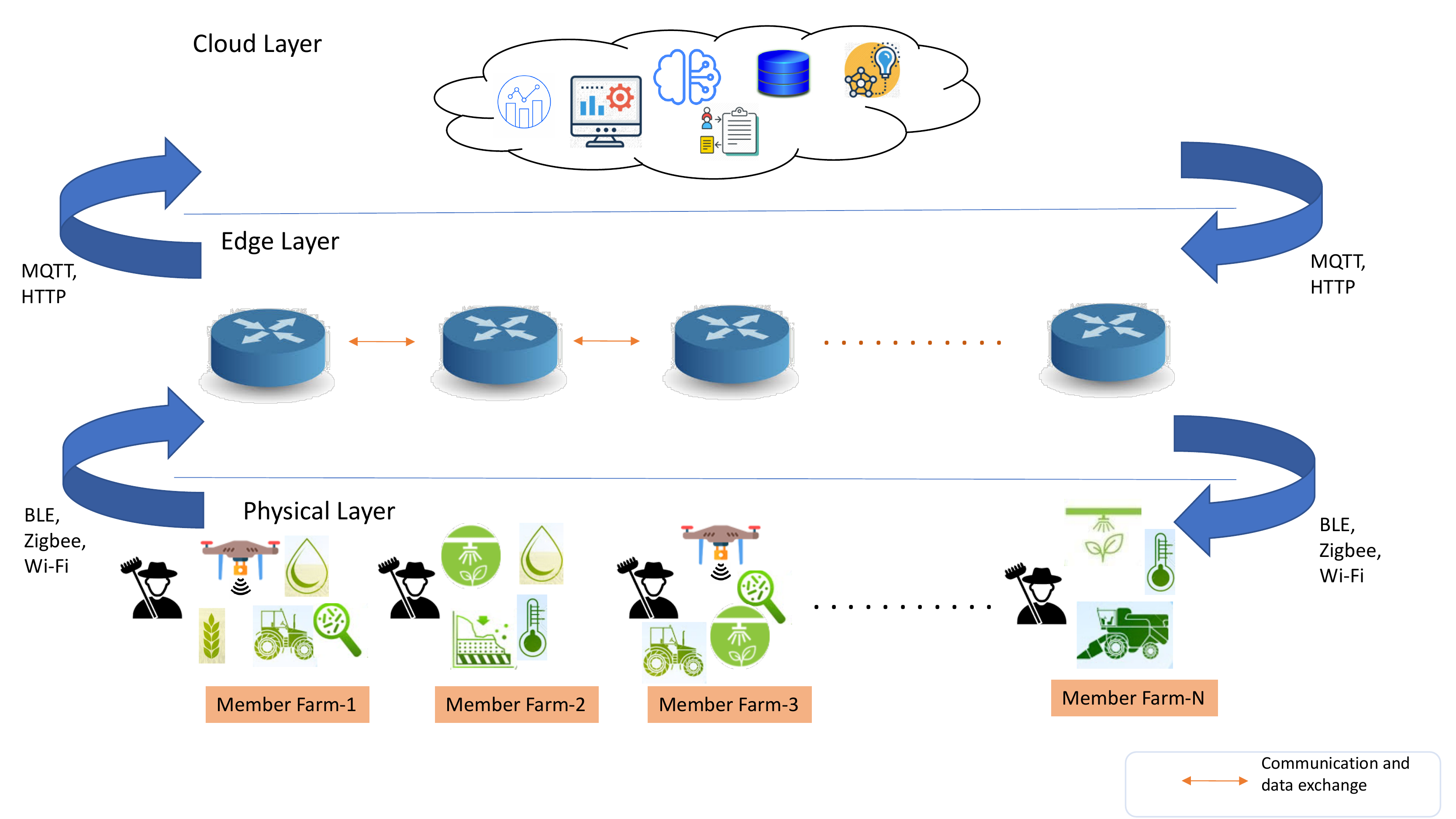}
\centering
\caption{Cooperative Smart Farming Architecture}
\label{fig:arch}
\end{figure*}
Such critical insights are only possible with the help of IoT devices that relay data from the farms in the cooperative hosted on cloud or edge servers. These farms reinforced with IoT devices come under the purview of smart farms which have been researched in many studies \cite{kamilaris2016agri, bacco2018smart}. Smart farms also employ remote sensing techniques to keep track of the covered land, monitor crop health and even classify weeds \cite{lottes2017uav}. 

Moreover, the use of technology in smart farms has been so far advanced as distinguishing between different plowing techniques and classifying them according to plowing depth \cite{tripicchio2015towards}. The use of Unmanned Aerial Vehicles (UAV) has been consistent in smart farms along with the use of IoT devices. An emerging setting for smart and cooperative farming is indoor farming, especially greenhouses, where IoT devices that share data with the cloud have been utilized to replicate the results achieved in outdoor farms \cite{jhuria2013image}.
 
In terms of implementation, there have been several instantiations of smart and connected farming ecosystems. These smart farms have been fitted with a variety of IoT devices that can interact with each other and also with the users to run different tasks. Similarly, REST-APIs have been used to encourage the cooperative development of applications and solutions to speed up adoption of the concept of smart and cooperative farming \cite{ryu2015design}. Data Modeling has been another important tool used in smart farming to perform a range of tasks, such as the very basic ones like feed modeling to more environmentally responsible ones as reducing methane emissions \cite{o2017modelling}. With many IoT devices in place, there is a sizable generation of continuous data streams in smart and cooperative farms. Big data technologies play an important role in the collection, storage and analysis of data and its conversion to meaningful information, which can in turn help in overall farm management and streamlining of farm processes \cite{wolfert2017big}.
 
With regard to the advances in smart farms, AI techniques are increasingly being used in the smart farming domain today. ML models and techniques too have begun to play a prominent role in enhancing the capabilities of smart farms. Image processing using artificial neural networks have been used to monitor plants both during growth and harvest periods for detecting diseases as well as for grading them \cite{jhuria2013image}. In addition to plants, ML has also been used to predict soil and land properties even the propensity of droughts to occur \cite{rezk2020efficient}. Also, with respect to autonomous operations in smart farms, models have been developed to minimize the instances of human intervention required for monitoring crops \cite{varghese2018affordable}. Another interesting application of AI capabilities in the smart farming context has been for weed classification where spectral images collected by small Micro Aerial Vehicles (MAVs) have been used to identify the weeds that may pose a danger to the surrounding plants or crops \cite{sa2017weednet}. 

\subsection{Security and Privacy issues on Cooperative Smart Farming}
The use of IoT devices in the smart farms has surely transformed the cooperative farming landscape, however it has also ushered in the risks attached to such devices. These risks translate to the various security and privacy risks that target IoT devices in general. Since a major part of the infrastructure of a cooperative smart farm is built around participating farms sharing data with each other, if a particular data source becomes corrupted, then it can have far reaching implications for all the farms in the bloc. Privacy by design is critical in such scenarios to protect the privacy sensitive information of any member farm from being subject to unwanted disclosure \cite{porambage2016quest}. Also, security of the data being circulated in CSF bloc is equally important, essentially with respect to the three triad - \textit{confidentiality, integrity,} and \textit{availability}. If this data is tampered with, then it can lead to devastating consequences such as wiping out entire crop harvest or withering of plants because of inefficient application of fertilizers. Therefore, it is crucial to ensure security and privacy be robust in smart farms.
 
Security and privacy issues in CSF architecture relate to authorization and trust, authentication and sever communication as well as compliance and regulations \cite{gupta2020security, gupta2020attribute}. Access control plays an important part in regulating the level of access to the many different IoT devices that may be present within CSF architecture. Federated access control reference model \cite{awaysheh2020next} is used to enhance the privacy of security of big data platforms. Similarly, there have been prior work on access control models for developing flexible access control models for securing access and data communication in IoT  \cite{ouaddah2017access, alshehri2017access, bhatt2017access1, bhatt2020abac}, and cloud-enabled IoT platforms \cite{bhatt2017access, gupta2020access, bhatt2019authorizations}. Data is transferred quite a bit among the cooperative smart farms as well as shared with the cloud servers responsible for aggregating the data for analysis. To combat these security threats there can be a number of solutions. A potential one solution is the application of Blockchain, with which the data transferred among the farms can be verified for its validity \cite{patil2017framework}, and another solution is based on trustworthy evaluation framework \cite{aladwan2020truste}. Also, with the increasing popularity of smart farms, another security challenge that arises is that these farms inadvertently play host to a plethora of security challenges \cite{barreto2018smart}. However, enforcement of data transfer agreements and verifying the approved learning received from the cloud servers is a very important task in the connected farm system. 
\begin{figure*}[t]
\centering
\includegraphics[width=0.8\textwidth]{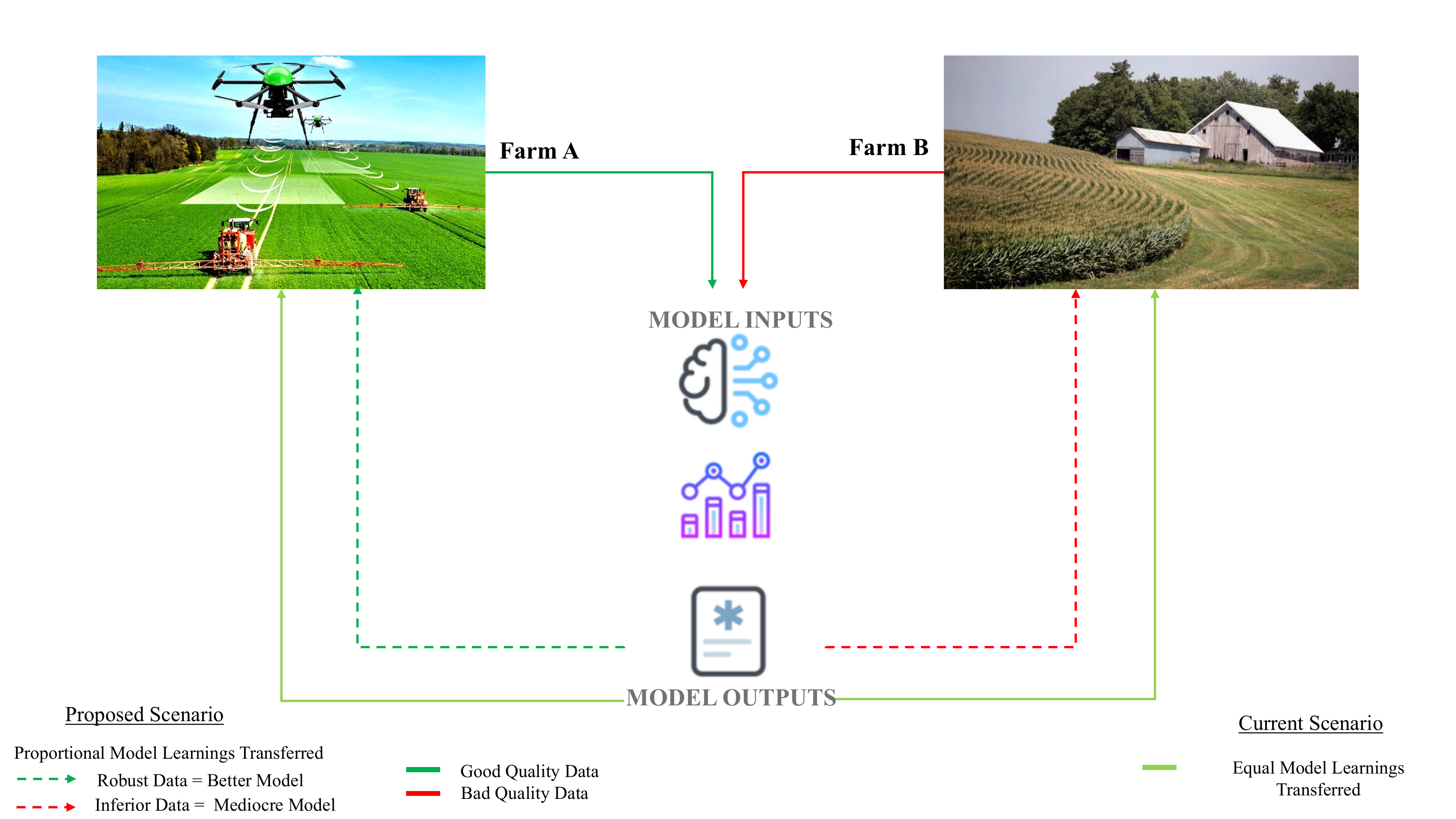}
\centering
\caption{Cooperative Smart Farming Model}
\label{fig:model}
\end{figure*}

Further, the issue of security and privacy in smart farming is not only restricted to data alone but encompasses both the human and technology components of CSF architecture. It is important that the farm owners are aware and trained on the latest threats that may pose a danger to their farms. On the other hand, the technology providers and vendors should provide end users with applications and software that is free from flaws and secure by design \cite{gundu2019iot}. To enhance the security in smart farms, researchers have also looked at using a private cloud infrastructure that combines data from heterogeneous sources to provide privacy secure data analytic capabilities to individual farms \cite{krintz2016smartfarm}. Such a solution can be ported into a collaborative mechanism that can effectively serve participating smart farms in a cooperative bloc. Ensuring proper device authentication of smart devices with low computing power, that are embedded in connected farm systems, can also act as an effective deterrent against cyber-attacks and minimize security violations \cite{chae2018enhanced}.

\subsection{Game Theory in Smart Farming}
This paper \cite{bruce2016game} investigates the utilization of game theory models for automated analysis of hyperspectral imagery data. The authors discuss two models on feature-level fusion and decision fusion for hypertemporal-hyperspectral datasets, and both models are implemented under the assumption where all players are rational. In agriculture, this research \cite{dillon1962applications} presents game theory model for decision making in risky situations. In beginning, a farmer makes plans for obtaining feasible goals and then simply carry out the plans. Later, he/she can face various uncertain issues in agriculture production. This study \cite{walker1959game}  provides possible solutions to actual decision problems of Iowa farmers. This paper \cite{cabrera2013application} deals with the application of game theory with solution to handle an agricultural economics problem. This study claims to establish Nash Equilibrium for an agricultural company that is considered together with its three sub-units. Gupta et al. \cite{gupta2020learner} presents a novel fair strategy for collaborative deep learning game, where all mobile edge devices \cite{gupta2020secure} choose any strategy to make own profit. The past studies show that no research has analyzed the farm's rational behavior in CSF. Therefore, we develop a game model for farms in CSF and analyze the game.

\section{Cooperative Smart Farming}
\label{usecase}
\begin{figure*}[t]
\centering
\includegraphics[width=.8\textwidth]{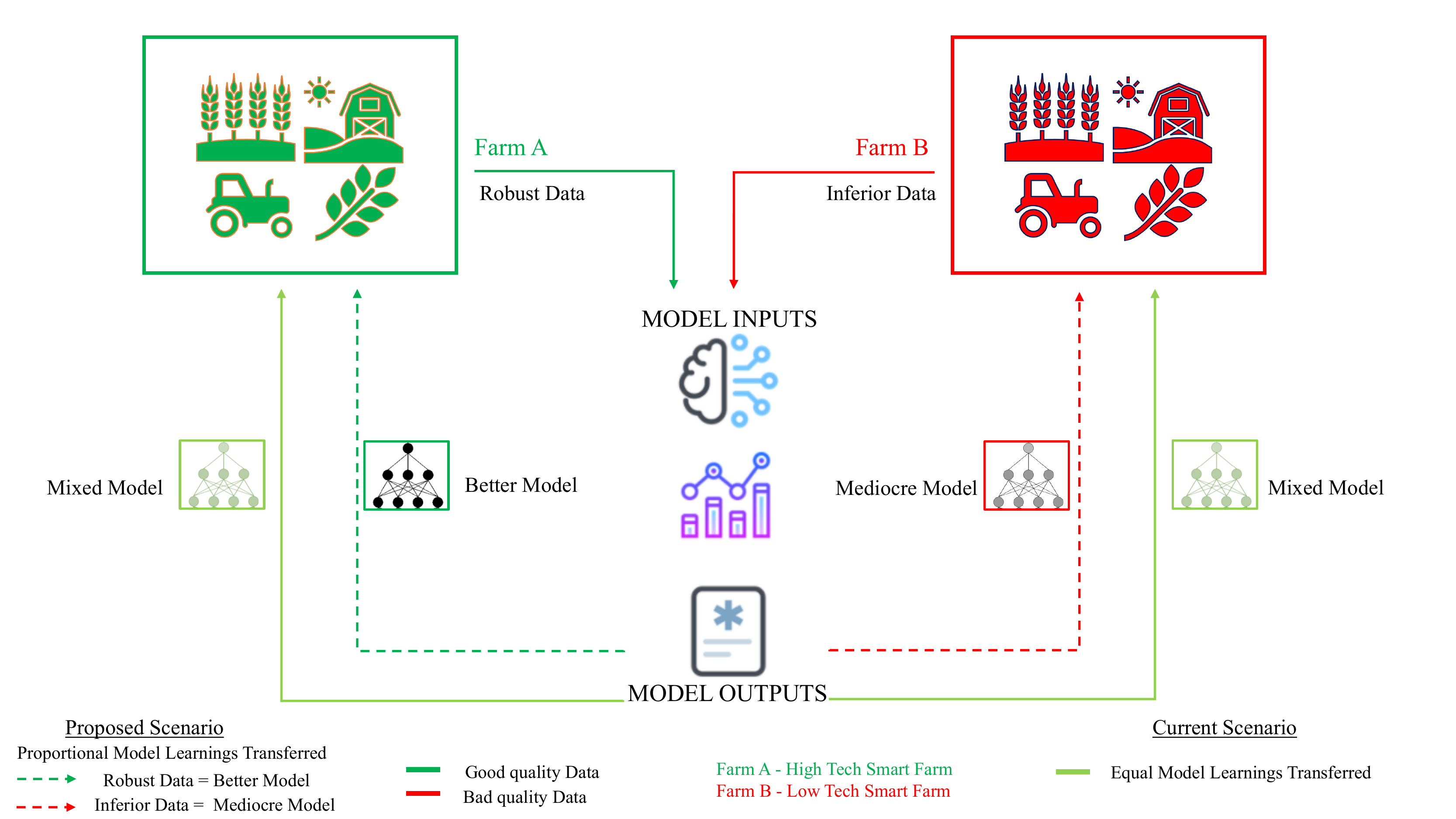}
\centering
\caption{Cooperative Smart Farm Model - Threat Scenario 1}
\label{fig:model-1}
\end{figure*}

Cooperatives are formal corporation which are financed, controlled, and owned by members for correlative advantage \cite{barton1989cooperative, berner2014successful, chukkapalli2020ontologies}. These cooperative (co-op) work is based on membership agreement, which has operational rules along with conditions and provide the benefit of shared resources. Such cooperatives operate in different sectors, such as grocery suppliers, water supplies, credit unions, utilities, farm suppliers, transportation and childcare. According to United States Department of Agriculture (USDA) \cite{usda}, the importance of co-ops is more meaningful in the rural communities. Figure \ref{fig:arch} presents the multi-layer architecture of CSF, where physical layer has Internet of Agriculture Things (IoAT) devices, the edge layer provides local real-time computation and analysis needed for smart resources, and cloud layer comprises of cloud services and co-op level.

\subsection{A Use Case}
We present a use case of multiple participating smart farms in a CSF bloc which is based on a game theory-based model. Participating smart farms are rational choice makers and will only become members of CSF bloc if they get tangible benefits from it. These benefits have to be more beneficial than would normally accrue to the farms if they operated alone based on their own capabilities. In this setting as shown in Figure \ref{fig:model}, CSF bloc provides its members, \textit{Farm A} and \textit{Farm B}, the benefits from a shared ML model. This model can provide numerous insights related to various factors, such as plant and soil health, soil and water quality, and crop output and yield management information, based on data collected from \textit{Farm A} and \textit{Farm B}. Therefore, if a smart farm wishes to use the information provided by the model, then it has to register and join CSF bloc so that it can avail of the learning from the model.
This ML model is trained on the cloud or edge server with data that is transferred to it from the participating farms. Generally, we assume that the larger the dataset collected from different farms the better ML model will be; thus, there will be an inherent incentive for the participating farms to become a member of CSF bloc and share their data in order to get better insights about their farming practices and in turn get more profits in the long run.

If a smart farm does not need the information being provided by the model, then it will not choose to join CSF bloc. For example, if a smart farm needs information about soil moisture or soil pH levels and the model only provides insights about water quality and yield management, then the farm will not participate in the bloc. Similarly, if earlier ML model in a bloc used to provide the soil moisture information, but going forward, due to a lack of good quality data the model stops providing that specific information, then a smart farm can choose to defect from CSF bloc.
Also, if a smart farm is technically superior and has abundant data and ML capabilities, then it will not join a CSF bloc. Instead it will operate alone, relying on its own capabilities. Similarly, if a participating smart farm acquires such superior capabilities while being a member of a CSF bloc then, it can choose to defect from the bloc. However, we envision that these scenarios are only fringe options that depends on different types of participating farms and may or may not be seen often in the context of our smart farms based CSF architecture.

\begin{figure*}[t]
\centering
\includegraphics[width=.8\textwidth]{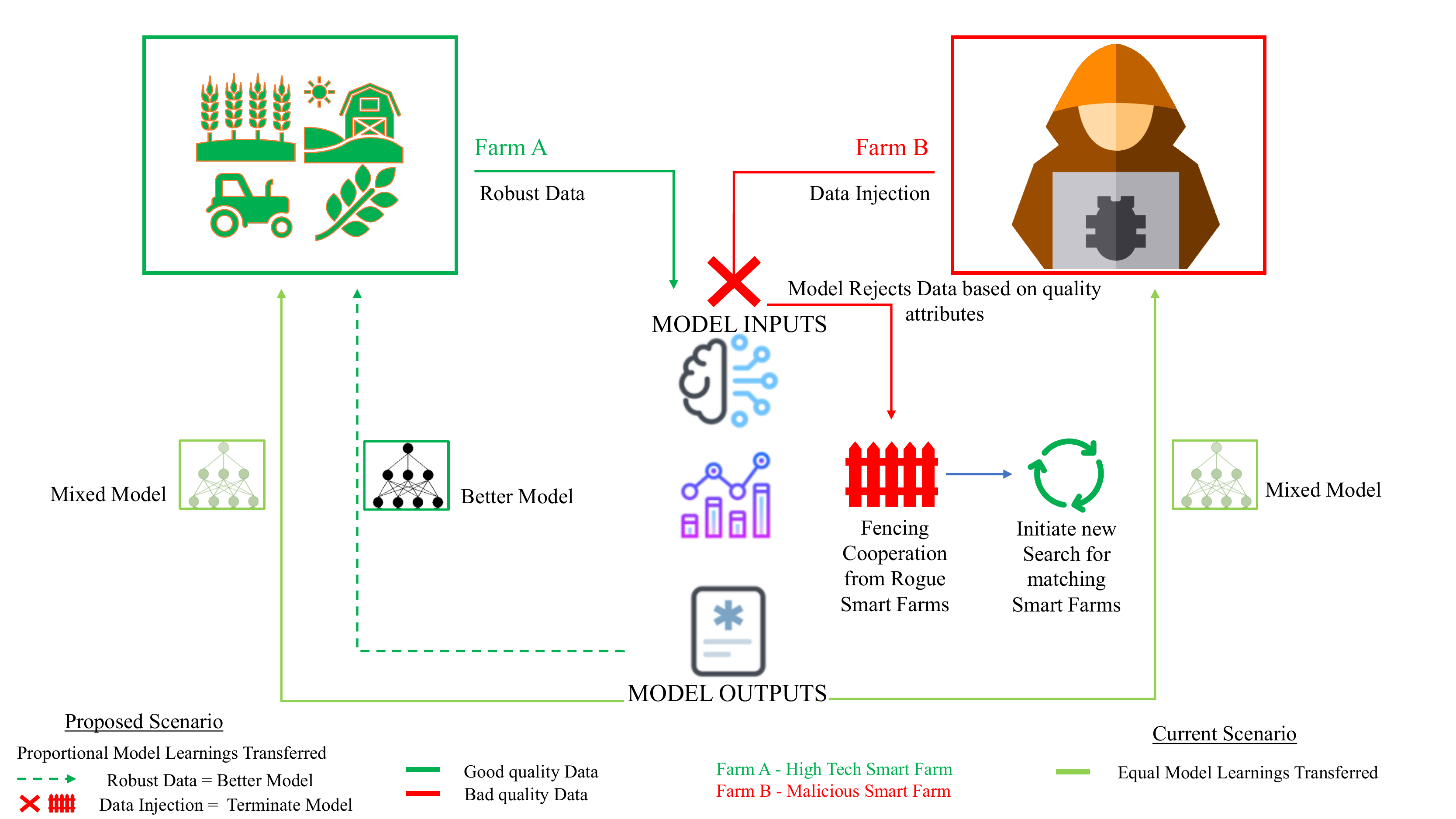}
\centering
\caption{Cooperative Smart Farm - Threat Scenario 2}
\label{fig:model-2}
\end{figure*}
\subsection{Threat Model}

\subsubsection{Scenario 1}
In a CSF bloc, there are two farms – \textit{Farm A} and \textit{Farm B}. \textit{Farm A} has various sensors, smart devices, robust data collection procedures in place, sophisticated instruments that collects high-quality data with regard to crop yield per hectare, soil moisture, water quality, irrigation schedules etc. \textit{Farm B} does not have such advanced sensors, devices, and methods to collect good quality data. Both farms share data with a cloud ML model with the instruments they have installed on their farms. Figure \ref{fig:model-1} shows this scenario. 

The data collection (what metrics are collected), storage (where are they storing and where – IoT Hub or Edge computing servers, or cloud) and sharing schedule (data is sent hourly, or daily or weekly) may be different or standardized for different participating farms. The cloud ML model uses data across different farms and trains model to make predictions regarding pesticide application, seeds needed per hectare, minimum irrigation levels needed to ensure good crop yield, predict scarcity of soil nutrients etc. The ML model shares the data with all the farms irrespective of their size or quality of the data contribution. In this scenario, \textit{Farm A} has good devices, data collected from those devices, and resources; however, \textit{Farm B} does not have good farm devices, data, and resources. Therefore, the ML model should process the data from \textit{Farms A} and \textit{Farm B} and then provide outputs/insights based on the quality of data that each farm shared rather than providing both with the same insights, as is the general case with the ML model learning. Figure \ref{fig:model-1} shows both current learning and insights sharing mechanism with solid green lines, and our proposed mechanism with dashed lines.

\subsubsection{Scenario 2}

Figure \ref{fig:model-2} depicts the second threat scenario where one of the farms could have malicious intents. Here, the scenario also consists of two farms, \textit{Farm A} and \textit{Farm B}, where \textit{Farm A} is a technologically advanced farm compared to \textit{Farm B} and \textit{Farm A} provides better quality data to the ML model, similar to scenario 1. Now, \textit{Farm B} decides to act rogue and stop sharing the entire data it collects, or does not replace outdated or broken IoT devices, or \textit{Farm B} has malicious intentions and does not collect good quality data. For instance, the malicious intent can either be competitors trying to sabotage a CSF and smart farms participating in that CSF, or data injection attacks from hackers for ruining crops for other farms and their ML model outputs. 
In scenario 1, \textit{Farm B} still gets the ML model’s prediction without actually cooperating and participating in the smart farming architecture. In scenario 2, \textit{Farm B} actively tries to sabotage the whole CSF infrastructure by compromising the data source from which the ML model is trained.
With our game theory-based fair strategy approach (described in later section) proposed in this paper, we try to solve this problem and ensure that the participating smart farms are equal and honest in all respects. As a consequence of providing high-quality and true data, the smart farms receive models and their outputs/insights that accurately represent the aggregate learning from CSF bloc.

\subsection{Rationality Assumption}
Prior research in smart farming \cite{sontowski2020cyber} have presented cyber-attacks on smart farming infrastructure where farmers or IoT devices are controlled by adversary. Malicious participants could arbitrarily deviate from suggested protocol in CSF or could arbitrarily drop communication between edge gateways and cloud. However, here we assume that farms are honest in a selfish environment. In the context of this paper, rationality means that a rational farm earns maximum profit by collaboration or individual modeling in CSF.

\section{Cooperative Smart Farming Game}
\label{gamemodel}


In this section, we present a game model of CSF with multiple member farms in selfish environment. This game model involves with $N$-players and is addressing the threat scenario 1 as discussed in \ref{usecase}. This game model refers as a cooperative smart farming game \textit{G}. In this game, IoT devices send their generated data to centralized cloud where data supports all the member farms through ML models. Co-op allows to aggregate only that data, which is generated from members of CSF and builds ML model for AI-assisted applications. This updated global ML model provide insight to all participating member farms via AI-assisted application, where exists a social dilemma for all defection behavior. 

Table 1 shows game approach between two players, one has high-quality data and other has low-quality data. List of symbols are shown in Table 2.

\subsection{Game Theoretic Model}
Game theory allows for modeling situations of conflict and for predicting the behavior of participants when they interact with each other. In CSF game \textit{G}, farms which sign a co-op agreement to be part of CSF are participants. They interact with each other through co-op cloud level without having any knowledge about each other. The Game \textit{G} is a static game, because all participants must choose their strategy. The Game \textit{G} is a tuple $(P,S,U)$, where $P$ is the set of players, $S$ is the set of strategies and $U$ is the set of payoff values.

\begin{itemize}
\item \textbf{Players} ($P$): The set of players $P=\sum_{i=1}^{N} P_{i}$ corresponds to the set of member farms which signed co-op agreement and allows to send generated data from their IoT devices through edge gateways to build ML model in CSF game \textit{G}. 

\item \textbf{Strategy} ($S$): Each participant $P_i$ can choose between two actions $s_i$ (i) Cooperative ($CP$) or (ii) Defective ($DF$). Hence the set of strategies in this game is $S$ = \{$CP$,$DF$\}. Strategy of each participant $P_i$ determines whether $P_i$ participates in CSF. In particular, if a participant $P_i$ plays $CP$ strategy, i.e., the member farm allows IoT devices to send the generated data to centralized cloud and extract model results via AI-assisted applications. In cooperative strategy, participant pays full cost. In contrast, if a participant $P_i$ neither sends the generated data through IoT devices to the cloud nor utilize the global ML model, i.e., the member farm $P_i$ plays $DF$ strategy. Thus, participants saves various costs including membership costs $c^o$, communication costs $c^m$, $c^m{}^{'}$ and storage cost $c^{s}$. Here, this participant is not part of CSF and builds ML model individually.

\item \textbf{Payoff} ($U$): The goal of each participants in CSF game \textit{G} is to maximize their utility, which is a function of the accuracy and its costs. In this game, we do not consider any malicious or threat activities from players; hence, the accuracy based on AI assisted application is benefit for each player. The players pay various costs including communication costs $c^{m}$, $c^{m'}$, storage cost $c^{s}$, and membership cost $c^{o}$, and also pay penalty $c^{p}$ if any player breaks co-op membership agreement. 
Total cost $c_i^t$ to participate in CSF can be characterized as
\begin{equation}
c_i^t = c^{o} + c^{p} + c^{m} + c^{m'} + c^{s} 
\end{equation}
\end{itemize}
\begin{table}
\centering
\caption{Game Table Between Two Players}
\label{tab:1}
\renewcommand{\arraystretch}{1}
\begin{tabular}{|p{1cm}|p{2.5cm}|p{2.5cm}|} \hline
\textbf{Game Options}  & \textbf{Low Data Farm} & \textbf{High Data Farm}\\\hline
\textbf{Low Data Farm} & Less robust model - Train model on (Low-Low) farms. & Cancel model training and move to find next pair of (High-High) farms. \\ \hline
\textbf{High Data Farm} &  Cancel model training and move to find next pair of (High-High) farms &  Robust model - Trained on matched sets on (High-High) farms. \\
\hline
\end{tabular}
\end{table}
\begin{table}[]
\def\arraystretch{1.15}
\caption{List of Symbols}

\begin{tabular}{c|l} \toprule
\textbf{Symbol}  &    \textbf{Definition}  \\
\bottomrule
$N$				 & Number of member farms\\
$n$& Total number of IoT devices at farm\\
$D\textsubscript{i}$ & Generated data from IoT device i\\
$P$ & Players \\
$S$ & Set of strategies\\
$U$ & Total payoff \\
$B$ & Coefficient \\
$a\textsuperscript{co-op}$ 	& Accuracy of model based on co-op level\\
$a$ 	& Accuracy of model without co-op \\
$c\textsuperscript{o}$ 	& Membership cost for co-op agreement\\
$c\textsuperscript{p}$ 	& Penalty to break co-op agreement\\
$c\textsuperscript{m}$ 	& Communication cost to send data to cloud\\

$c\textsuperscript{m'}$ 	& Communication cost to download updated model from cloud\\
$c\textsuperscript{s}$ 	& Storage cost to store data on cloud\\
$c^{plocal}$          & Computation cost to build model locally \\
$c_i\textsuperscript{t}$ 	& Total cost for participating in cooperative smart farming  \\

$C_i$					& Number of cooperative participants\\

$N-C_i$					& Number of defective participants\\

\bottomrule
\end{tabular}

\label{tab:Symbols}

\vspace{2mm}

\end{table}

\begin{figure}[t]
\centering
\includegraphics[width=0.45\textwidth]{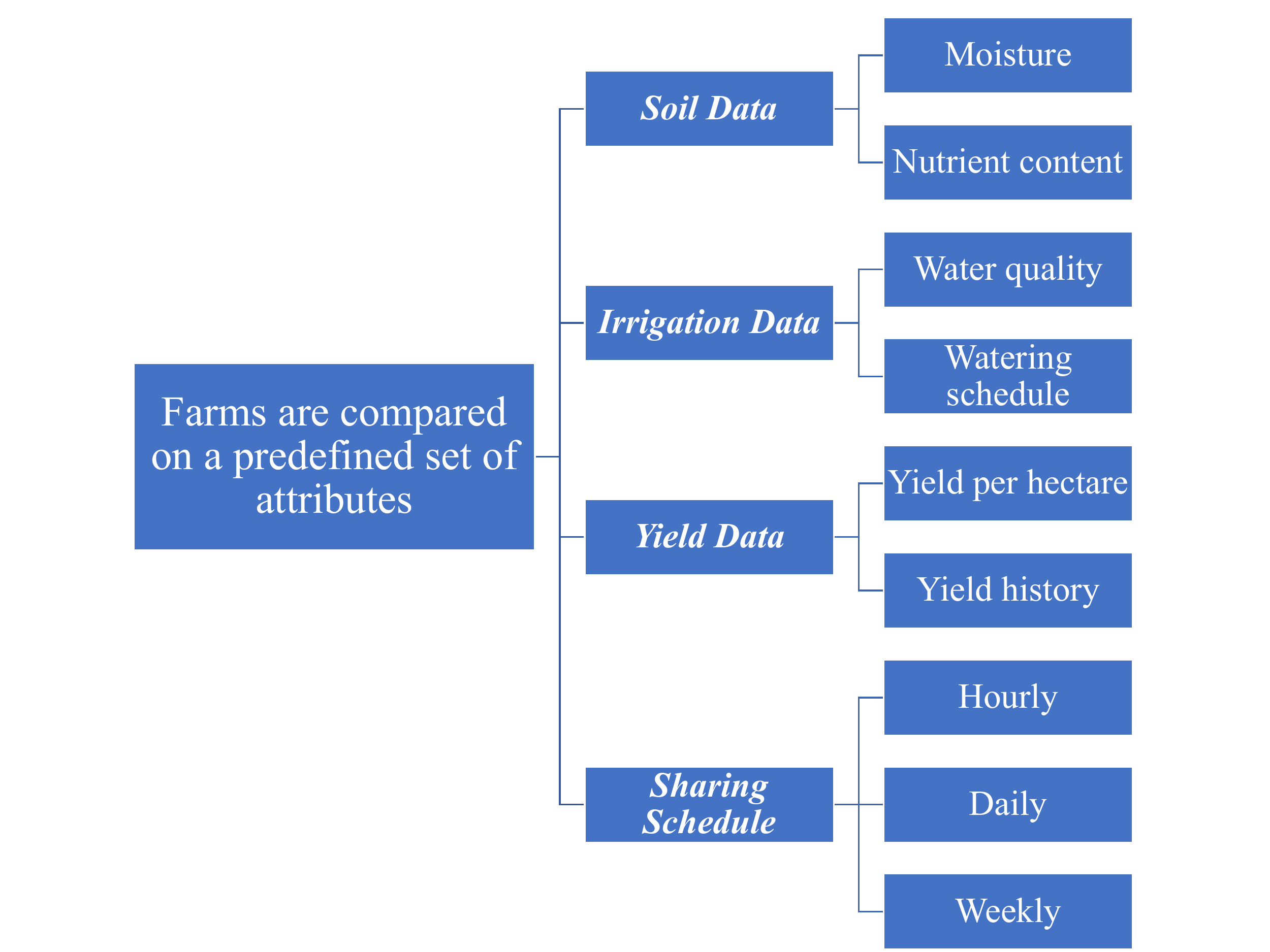}
\centering
\caption{A Sample of Model Attributes}
\label{fig:attri}
\end{figure}

\begin{figure*}[t]
\centering
\includegraphics[width=0.8\textwidth]{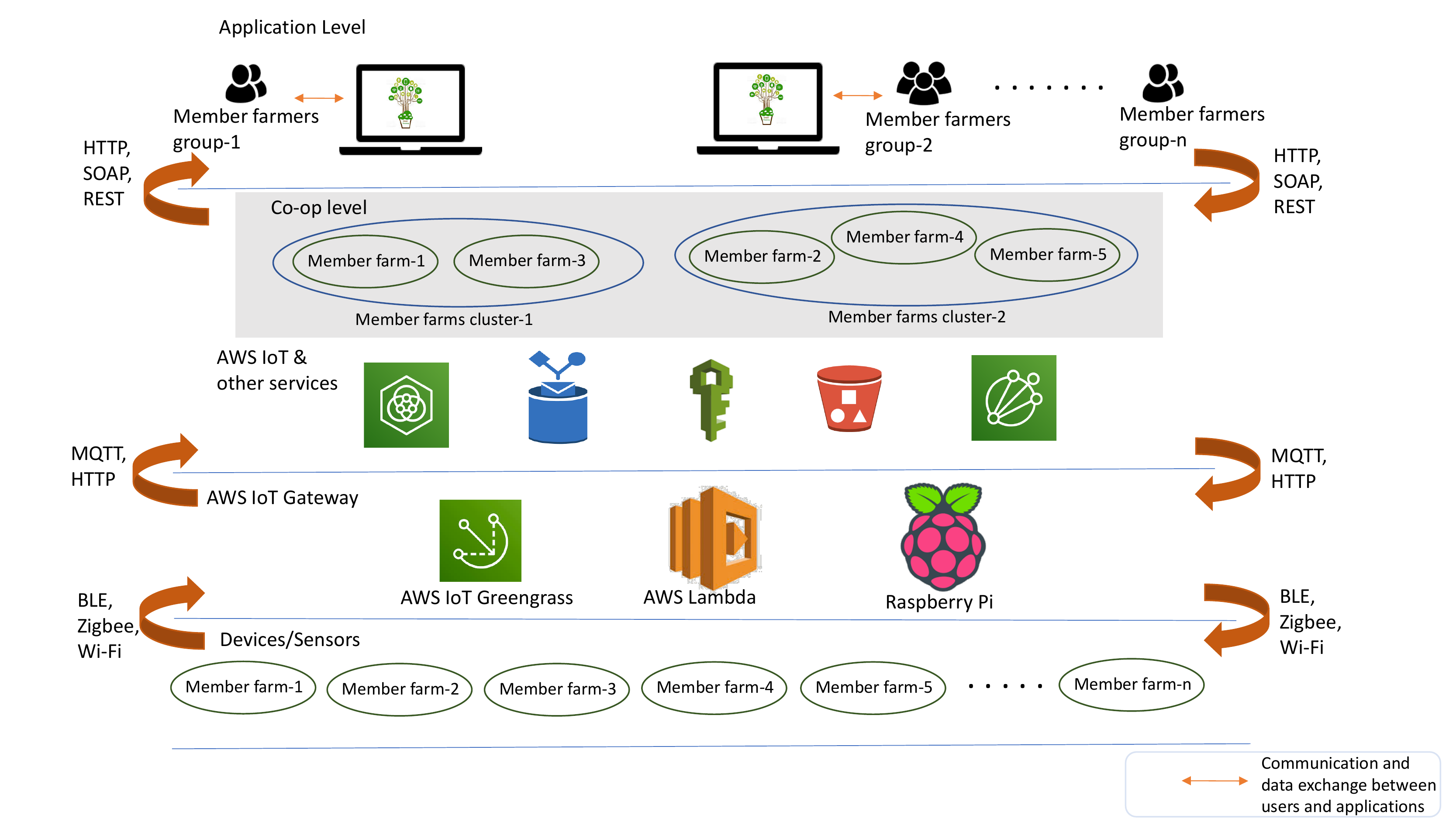}
\centering
\caption{Proposed Implementation of Cooperative Smart Farming (CSFs) in AWS cloud}
\label{fig:implementaion}
\end{figure*}

Here, the benefit and the cost are not on the same scale as the first depends on the accuracy of ML model while the latter on different types of associated costs. To make them comparable, we introduce a coefficient: the benefit is multiplied with B.

Now, we compute the payoff of player $P_i$ in this game. If we assume that the player $P_i$ is cooperative, i.e., $P_i$ $\in$ $CP$. Similarly, if $P_i$ is defective, i.e., $P_i$ $\in$ $DF$, and these payoffs can be defined as follows.

\begin{equation}
u_i(CP) = B(a\textsuperscript{co-op}) - (c^{o} + c^{p} + c^m + c^{m'} + c^{s})
\end{equation}

\begin{equation}
u_i(DF) = B(a) - (c^{plocal})  
\end{equation}

Based on the above calculated utilities, we analyze the game G as discussed in the game analysis. 

The most fundamental game-theoretic concept, Nash Equilibrium, which is introduced by John Nash \cite{nash1951non} is used to solve strategic behavior of participants.

\textbf{Definition 1.} A Nash Equilibrium is a concept of game theory where none of the players can unilaterally change their strategy to increase their payoff. 

In a non-cooperative game, players do not have any motivation to deviate unilaterally from the given strategy. The prisoner's dilemma is a standard example to analyze a game and shows that two rational players cooperate, come out with best outcome, and the players do not cooperate with one another, then they choose defecting strategy in the hope of attaining individual gain at the rival’s expense. In prisoners’ dilemma defecting strategy strictly dominates the cooperation strategy. Hence, the only Nash Equilibrium in prisoners’ dilemma, is a mutual defection. 

Based on the cost and benefit of player to build ML model, we build a CSF game model \textit{G}. In the following theorems, we show that the CSF game \textit{G} is a public good game. 

\textbf{Theorem 1.} \textit{ In cooperative smart farming game G, if each participant builds its local machine learning model individually, then we can establish All-Defective strategy profile as a Nash Equilibrium.}

\textit{Proof.} Let us consider all \textit{N} participants follow defective-$DF$ strategy where all participants neither send their generated data from IoT devices through gateways to cloud, nor upload any updated ML model/extract any results from cloud, i.e., no membership fees $c^{o}$, no communication between IoT devices and cloud. So, participants do not pay any communication costs $c^m$, $c^{m'}$, and storage cost $c^{s}$. Now each participant $P_i$ trains local data sets $D_i$ to build its ML model individually and pays only computation cost $c^{plocal}$. None of participants cannot change his strategy profile unilaterally. Let us assume if a participant deviates from defect-$DF$ strategy to cooperate-$CP$ strategy unilaterally, then participant will pay some costs ($c^o$+ $c^m$ + $c^{m'}$ +$c^s$ + $c^{p}$), which is greater than $c^{plocal}$. The payoff of cooperate-$CP$ strategy is less than defect-$DF$ strategy, so All-$DF$ is a Nash equilibrium profile and G is a public good game. 

Theorem 2 shows that we can never enforce an All-Cooperative $CP$ strategy in game G, and therefore, we could not establish a Nash Equilibrium.

\textbf{Theorem 2.}\textit{ In cooperative smart farming game G, if each participant builds its local ML model, then we cannot establish All-Cooperative strategy profile as a Nash Equilibrium.}

\textit{Proof.} We first assume that all \textit{N} participants are members of CSF (i.e., All-cooperative $CP$ strategy profile) and pays mandatory costs (membership cost, communication costs, and storage cost) and optional cost (penalty cost). We can compute the payoff of each participant $P_i$ by Equation (2). Hence, if a participant will deviate from the cooperation and play defection unilaterally, its payoff would be equal to Equation (3), which is always greater than cooperative payoff. Hence, each participant has incentive to deviate unilaterally and increases its payoff. Then, the All cooperate-$CP$ strategy profile is never a Nash Equilibrium.

\section{Proposed Approach - A Fair Strategy}
\label{approach}

We present the proposed model to ensure member farms in CSF can be maintained with a fair-game strategy.  Figure \ref{fig:model-1} and Figure \ref{fig:model-2} present CSF models of use case threat scenario 1 and scenario 2, which we discussed in section \ref{usecase}. Our goal is to design a mechanism for eliciting cooperation in CSF and solves the issue discussed in scenario 1 where one farm (Farm A) has advanced and good quality devices and provide good quality data. The proposed fair strategy enforces member farms for cooperation in CSF; however, Theorem 1 and 2 proves that member farm defect from CSF game, i.e., after some number of iterations, member farm will not be part a of CSF if they will not get any benefit from other member farms.
The quality of each dataset of each participant is calculated based on data attributes at co-op level. The generated data is classified using Support Vector Machine (SVM) algorithm and a result of sample data attributes is shown in Figure \ref{fig:attri}. This shows different categories of data items and hierarchy of attributes - \textit{soil data, irrigation data, yield data}, and \textit{sharing schedule} for data and resources. Each of these categories are further broken down into other attributes that provide detailed information about these attributes. For instance, soil data include moisture and nutrient content, irrigation data include water quality and watering schedule, etc. Based on such identified attributes for a ML model, we can define the data being collected and transmitted by smart farms to the CSF is quality data or not. 

As shown in Table 1, the best strategy is to classify the smart farms based on the quality of data they provide to the model and co-op, such as high-quality data farms should participate in high-quality CSF, and low-quality data farms can be grouped together and should participate in low-quality CSF. We believe that this strategy will allow the smart farms to co-operate in CSFs and take maximum benefit from it, and can reduce the problem of defective farms. However, further research on data attributes for ML models to efficiently classify the smart farms is necessary. We plan to focus on this aspect in our future work.
In the context of the CSF game \textit{G}, before the start of the game, each participant has to choose his strategy between $CP$ and $DG$ to play the CSF game \textit{G}. However, in the beginning of this game, each participant is in dilemma to choose strategy, which can depend on other participant's strategy. In CSF, each member farm does not know about the quality and quantity of other participant's data. Here, we propose a novel fair strategy employing K-means clustering at co-op cloud level. K-means clustering is an unsupervised ML technique, whose purpose is to segment a data set into K clusters. The co-op cloud will create clusters based on quality and quantity of their dataset to enforce in CSF game \textit{G}. Algorithm 1 shows our proposed strategy using data attributes (which would be defined for qualitative and quantitative data by co-ops) and K-means clustering. 
Using this algorithm and classified dataset, we can generate sustainable CSFs without the issue defective farms.
\begin{algorithm}[H]
\SetAlgoLined
\caption{Proposed Fair Strategy}
\begin{algorithmic}[1]
\STATE{Each participants send data $D_i$ to co-op cloud level.}
\STATE{The data is classified based on data attributes using Support Vector Machine (SVM) algorithm.}
\STATE{The classified data is tested on build-in ML model to calculate accuracy of each data set.}
\STATE{Apply k-means clustering algorithm on accuracy value of each participant i.}
\STATE{create the clusters at co-op level.}
\IF{participant i belongs a cluster with at least one other participant j}
\STATE $P_i$ , $P_j$ $\in$ $CP$
\ELSE
\STATE $P_i$ $\in$ $DF$
\ENDIF
\end{algorithmic}
\end{algorithm}

\section{Proposed Implementation}
\label{result}
In this section, we present a proposed implementation for CSF using proposed fair strategy utilizing AWS cloud and IoT platform, shown in Figure \ref{fig:implementaion}. In CSF, various smart member farms collaborate with each other. As shown in Figure \ref{fig:implementaion}, the data is collected from different smart member farms, which need to be classified using SVM algorithm. This proposed implementation utilizes AWS IoT core\footnote{https://docs.aws.amazon.com/iot/latest/developerguide/what-is-aws-iot.html} in cloud, AWS Greengrass\footnote{https://aws.amazon.com/greengrass/} running on edge gateways (e.g., Raspberry Pi enabled gateway) that can enable edge computing as required, and IoT devices/things corresponding to specific use cases requirements. 
The edge gateway can be enabled using Raspberry Pi\footnote{https://www.raspberrypi.org/} that host the AWS Greengrass deployment, and customized lambda\footnote{https://aws.amazon.com/lambda/} functions that can run on the gateway for edge communication and computation, sending notifications, or enforcing access control and privacy policies. AWS Greengrass allows farmers to securely connect their smart devices at the edge of the network to gather data. When connectivity is re-established, the data will synchronize with AWS cloud and its IoT services using MQTT/HTTP protocols.

For enabling a secure architecture, AWS IoT core provides mutual authentication based on certificates and encryption for secure data transfer and also enable different access control levels, which have been discussed in \cite{bhatt2017access, gupta2020access}. The data can be stored at AWS cloud storage and be analyzed at the co-op level for classified member farms based on the quality and quantity of their data attributes.  The member farms cluster will be created using proposed fair strategy at co-op cloud level, then the member farms will be part of CSF. AWS IoT analytic service can be used to analyze the data using specific ML models and provide meaningful insights to the member smart farms through the smart farm applications as shown in the top layer of the proposed implementation architecture.

\section{Conclusion}
\label{conclusion}
In this paper, we present a system model of cooperative smart farm (CSF) and introduce the problem of strategic behavior of farmers in CSF. We analyze rational behavior of member farms in CSF using game theory model. We also analyze the Nash Equilibrium strategy profile for each scenario, where member farms are enforced to cooperate based on some data attributes identified by CSFs and using our proposed fair strategy in CSF. This work presents an extensive understanding of non-cooperative behavior of member farms in CSF. For future work, we plan to do further research on data attributes identification for ML models that can be used by CSFs to better classify the smart farms based on the quality and quantity of data. We also plan to implement the model with farming data set and evaluate the accuracy of ML model using proposed fair strategy. 

\balance
{
\bibliographystyle{IEEEtran}
\bibliography{References}
}
\end{document}